\documentclass[%
 apl,
 amsmath,amssymb,%
reprint,%
superscriptaddress,
]{revtex4-1}
\usepackage{array}
\usepackage{graphicx}% Include figure files
\usepackage{dcolumn}% Align table columns on decimal point
\usepackage{bm}% bold math
\usepackage{cleveref}
\usepackage{siunitx}
\usepackage{xcolor}

\newcommand{\FeO}{$\alpha$-Fe$_{2}$O$_{3}$}

\newcommand{\approptoinn}[2]{\mathrel{\vcenter{
  \offinterlineskip\halign{\hfil$##$\cr
    #1\propto\cr\noalign{\kern2pt}#1\sim\cr\noalign{\kern-2pt}}}}}

\renewcommand{\vec}[1]{\mathbf{#1}}

\usepackage{lineno}

\graphicspath{{/Users/radaelli/Documents/Toshiba_Labtop/synch/Projects/Fe2O3/Jacopo_Paper_2019/Up_to_Date_Graphics/}}

\begin{document}
%\linenumbers
%\preprint{}

\title[Title]{Micromagnetic modelling and imaging of vortex$\vert$merons structures in an oxide$\vert$metal heterostructure}

\author{P. G. Radaelli}
\affiliation{Clarendon Laboratory, Department of Physics, University of Oxford, Parks Road, Oxford OX1 3PU, United Kingdom}
\author{J. Radaelli}
\affiliation{Clarendon Laboratory, Department of Physics, University of Oxford, Parks Road, Oxford OX1 3PU, United Kingdom}
\affiliation{ Department of Physics, University of Durham, Lower Mountjoy, South Rd, Durham DH1 3LE, United Kingdom}
\author{N. Waterfield-Price}
\affiliation{Clarendon Laboratory, Department of Physics, University of Oxford, Parks Road, Oxford OX1 3PU, United Kingdom}
\author{R. D. Johnson}
\affiliation{Clarendon Laboratory, Department of Physics, University of Oxford, Parks Road, Oxford OX1 3PU, United Kingdom}
\affiliation{Department of Physics and Astronomy, University College London, London, WC1E 6BT, United Kingdom}

\date{\today}

\begin{abstract}
Using micromagnetic simulations, we have modelled the formation of imprinted merons and anti-merons in cobalt overlayers of different thickness  (1-8 \si{\nano\meter}), stabilised by interfacial exchange with antiferromagnetic vortices in \FeO.  Structures similar to those observed experimentally could be obtained with reasonable exchange parameters, also in the presence of surface roughness.  We produce simulated meron/antimeron images by magnetic force microscopy (MFM) and nitrogen-vacancy (N-V) centre microscopy, and established signatures of these topological structures in different experimental configurations. 
\end{abstract}

%\pacs{75.85.+t,}% PACS, the Physics and Astronomy
\maketitle

\section{Introduction}

`Oxide electronics' aims at combining the multifunctional properties of transition metal oxides with more traditional spintronic approaches and represents one of the most promising pathways to post-CMOS computing \cite{Manipatruni2018}.  One approach is to exploit the rich real-space topological properties of oxide domains to create structures such as vortices and skyrmions, which could be `imprinted' onto ferromagnetic (FM) read-out overlayers.  Heterostructures of this kind, particularly those built with rare-earth-free materials, could be used as  high-density, non-volatile memories with a high degree of thermal stability.

Skyrmions are the best known example of magnetic topological object, and have received an enormous amount of attention (see for example \cite{Jiang2017, Lancaster2019} for recent reviews).   Magnetic merons/antimerons (essentially flat vortices/anti-vortices with an out-of-plane core) have been known to exist as closure domains in magnetic nano-dots since the early 2000's \cite{Shinjo2000}, and have more recently been observed in extended systems, either as intermediate stages of skyrmion array formation in chiral magnets \cite{Yu2018} or as light-induced metastable magnetic textures in the absence of in-built chirality \cite{Eggebrecht2017}. Both skyrmions and merons can be thought of as projections onto a tangent plane of a vector field defined on the surface of a sphere, the projection point being either the centre of the sphere (meron) or the opposite pole (skryrmion) \cite{Lancaster2019}.    As such, these objects can be characterised by a so-called \emph{topological charge} or \emph{winding number}, $w$.  The magnitude $|w|$ counts how many times the vector field wraps around the sphere or half-sphere, while the sign of $w$ defines the direction of wrapping.  The topological charge, which can be calculated as a surface integral of the projected field \cite{Lancaster2019}, is a positive (negative) integer for  skyrmions (anti-skyrmions) and a positive (negative) half-integer for merons (anti-merons) \cite{Lancaster2019}.   One important property of the topological charge is that, being an integer or half-integer, it must change discontinuously, and is therefore  \emph{left invariant by a smoothly-varying rotations in spin space}.  In other words, in order to alter the topological charge of a given object, one must introduce a singularity in the local field.  Since this tends to be associated with a high energy cost, these objects are often said to be `topologically protected'.  In real magnetic systems, topological object do not enjoy an absolute protection (for example, they can annihilate with their own anti-particles), but are often very stable against thermal fluctuations. 

Using a combination of X-ray linear/circular dichroism photoelectron emission microscopy (XMLD/XMCD-PEEM), we have recently demonstrated that \emph{antiferromagnetic} (AFM) planar vortices and anti-vortices exist in \FeO, and that these structures are `imprinted' as FM vortices onto a 1 \si{\nano\meter} soft Co overlayer \cite{Chmiel2018}.  Although our XMCD-PEEM measurements were not conclusive due to limitations in spatial resolution, they were consistent with an out-of-plane component of the Co spins, which would make the Co structures merons/anti-merons \cite{Callan1978} rather than planar vortices.  If corroborated, the observation of merons/anti-merons would be extremely important, since the out-of-plane spin component would represent a convenient 2-bit state, which, similar to skyrmions, is to a large extent topologically protected \footnote{Strictly speaking, a spin-up meron can be converted to a spin-down antimeron by a global rotation in spin space, but this requires to rotate a large number of spins by wide angles.}.   Another conclusion of our experimental work was that spins in Co are co-aligned with the \FeO\ AFM spins, indicating that the interaction responsible for the vortex$\vert$meron coupling is akin to exchange bias \cite{Nolting2000} (hereafter, we refer to this as `exchange-bias interaction') rather than the 90-degree interaction observed in other systems \cite{Papp2015}.  This raises another important question: since AFM spins in \FeO\ have opposite directions for adjacent terminations, how can Co merons/anti-merons be stable in the presence of surface roughness?

In this paper, we model the combined vortex$\vert$meron  and anti-vortex$\vert$anti-meron structures we have observed in \FeO$\vert$Co using micromagnetic simulations.  We demonstrate that Co merons/anti-merons are stabilised by an underlying  \FeO\ vortex/anti-vortex due to a competition between exchange stiffness, exchange-bias and magnetostatic interactions.  Correspondingly, the scale of the Co features is governed by the two exchange lengths, $L_{ex.b}$, which accounts for the field induced by the interface, and the usual magnetostatic length, $L_{ms}$.  We also determine the scaling of the meron core with the exchange parameters and film thickness, and establish that (anti-)vortex$\vert$meron structures are stable for rough interfaces, provided that the characteristic scale of the roughness is less than the exchange lengths.  Finally, we construct simulated scanning probe microscopy (SPM) images of the Co features using both magnetic force (MFM) and nitrogen-vacancy (N-V) centre  microscopy, and identified characteristic signatures that could be detected in the experiments.   Although our calculations and simulations are carried out for the \FeO$\vert$Co system, our methodology is  of general validity for exchange-coupled topological structures, and could be applied to a variety of systems of interest for oxide electronics and spintronics.  Light-induced metastable vortices \cite{Eggebrecht2017} are also described by our analysis as a limiting case in which $L_{ex.b}=0$.

The paper is organised as follows: in section \ref{section: Analytical considerations} we make dimensional considerations based on the key physical parameters and discuss a simple analytical model of a vortex$\vert$meron structure.  In section \ref{section: Micro-magnetic modelling}, we discuss our approach to micromagnetic simulations, in particular, providing a conversion between the atomic-scale and micromagnetic parameters, simulating surface roughness and constructing simulated SPM images.  Section \ref{sec: Results} contains the main results concerning meron stability (also in the presence of surface roughness), and the scaling of the meron/anti-meron cores, as well as our simulated MFM and N-VM images, and is followed by a short conclusion.

\section{Theory}
\label{section: Analytical considerations}

\subsection{Brief description of the physical system}
Our goal was to model the coupled vortex$\vert$meron structures observed by XMLD/XMCD-PEEM at room temperature (RT) (see ref. \cite{Chmiel2018}).  In this experiment, the physical system consisted of a 10 \si{\nano\meter} epitaxial [001] \FeO\ film grown on sapphire (Al$_2$O$_3$), with 1 \si{\nano\meter} FM Co capping layer grown at RT by DC sputtering.  The RT magnetic structure of  \FeO\ consists of collinear AFM layers (we ignore the small in-plane spin canting), stacked along the [001] direction in a repeated pattern `+ -  - +'.  All spins are perpendicular to the stacking direction and are aligned along one of the symmetry-equivalent \{100\} directions, so that six equivalent domains are possible.  A network of AFM vortices/anti-vortices are experimentally observed by XMLD-PEEM, where six domain meet at a single point.  Very similar topological structures are also observed by XMCD-PEEM in the Co overlayer exactly on top of the \FeO\ structures and having the same vorticity (vortex/antivortex character and direction of rotation).  The XMCD-PEEM vector-map intensity (proportional to the in-plane projection of the magnetic moment) shows a pronounced dip near the FM vortex cores, suggesting the presence of an out-of-plane ($z$) component, which is characteristic of \emph{merons}.  Since the size of the observed meron `cores' (i.e., the region where a sizeable $z$ component exists) was comparable to the typical X-PEEM resolution of 20-50 \si{\nano\meter},  it was not possible to establish the actual core size with any confidence.

\subsection{Feature sizes: dimensional considerations}
\label{sec: features size}

Consistent with our experimental findings, we will assume that Co experiences a bulk FM self-interaction, described by an exchange stiffness $A_{ex}^{\rm{Co-Co}}$ (in \si{\joule\per\meter}), as well as a surface interaction with \FeO\, described by an `exchange bias' constant $K_{ex.b}^{\rm{Fe-Co}}$, having dimensions \si{\joule\per\square\meter} (we will drop the unambiguous superscripts in the remainder).

In our simulations, we will assume that the domain structures in \FeO\ are rigid (i.e., not affected by the presence of the overlayer), that they have much narrower domain walls than those in Co and that all the \FeO\ spins lie in plane. At present, there is no experimental verification for these assumptions, which may not in fact be entirely correct.  In fact, since the energies of the in-plane and out-of-plane spin orientations are finely balanced, \FeO\ could even support AFM merons with an out-of-plane core,\cite{Galkina2010} \footnote{We have in fact obtained very recent experimental evidence of the existence of out-of-plane meron cores in pure \FeO.} while `reverse imprint' of a FM overlayer on an AFM has been previously discussed for other materials.\cite{Scholl2004}.  However, such a coupled problem would be intractable at the micromagnetic level, while the effect of a finite width of the \FeO\ domain walls can be easily included in our models, should any solid experimental evidence emerge.  We therefore believe that our assumptions are justified, in that they provide a simplified but useful model of the relevant physics.

Initially, we we will also assume that \FeO\ has a FM termination with no roughness (we will relax this assumption later).  With these assumptions, the other key physical parameter in the problem is the thickness $d$ of the Co film.  From these parameters, one can construct a length:

\begin{equation}
\label{eq: ex_l_def}
L_{ex.b}=\sqrt{\frac{A_{ex}d}{K_{ex.b}}}
\end{equation}

There is also a second length-scale in Co, the `conventional' magneto-static length \cite{Abo2013}, unrelated to the presence of \FeO\ and given by:

\begin{equation}
L_{ms}= \sqrt{\frac{2 A_{ex}}{\mu_0M^2}}
\end{equation}

where $M$ is the Co magnetisation.  From this simple analysis one should conclude that the size of any magnetic feature in Co should be determined by the competition between two lengths, $L_{ex.b}$ and $L_{ms}$, which control the `surface' and `bulk' physics of the problem, respectively.   Moreover, when one of the lengths is much larger than the other, Co features should roughly scale with the \emph{smaller} of the two lengths.

One could test this prediction by calculating, for example, the shape and width of a N\'eel domain wall induced by the presence of a sharp anti-phase boundary in the underlying AFM material.  Problems of this kind have been discussed at the micromagnetic level since the sixties \cite{Aharoni1966}, and involve differential equations imposing zero torque on each magnetisation element \footnote{The general form of the differential equation for a 180\si{\degree} domain wall is $\Theta''(x) -\frac{s(x)}{L_{ex.b}^2 } \sin \Theta(x)=0$, where $\Theta(x)$ is the spin angle of the equilibrium structure at position $x$ , $L_{ex.b}$ is the exchange bias length and $s(x)=\text{sgn}(x)$ is the sign function.  The second derivative of $\Theta(x)$ is clearly discontinuous at the origin, but a closed-form solution of this equation exist and can be expressed in terms of Jacobi functions.  The linear domain wall is a reasonable approximation of these solutions.}.  Although finding full analytical solutions is beyond the scope of this paper (which focusses on numerical solutions of these equations when $L_{ex.b}$ and $L_{ms}$ are comparable), in  Appendices \ref{app: linear_antiphase} and \ref{sec: meron_calculations} we present simple solutions of the linear domain wall and of the vortex-meron problem, respectively, \emph{ assuming} the shape of these structures to be a known function.  The linear N\'eel domain wall problem with uniform spin rotation is very similar to the well-known calculation of the width of a Bloch domain wall in the presence of magnetocrystalline anisotropy \cite{Abo2013}, and leads to a very similar result:  

\begin{equation}
\label{eq: domain wall width}
W_N=\pi \left(1-\frac{2}{\pi}\right)^{-1/2} \! \! \! L_{ex.b} \approx 5.21  \, L_{ex.b}
\end{equation}

where $L_{ex.b}$ replaces the usual magnetocrystalline anisotropy exchange length \cite{Abo2013}.  Eq. \ref{eq: domain wall width} is strictly valid in the limit $\frac{d}{L_{ex.b}} \ll 1$.   This is appropriate throughout most of the range we consider, since we typically set $d=$ 1 \si{\nano\meter} in agreement with \cite{Chmiel2018}, while typical domain wall widths in our simulations are $\ge 4$  \si{\nano\meter}.  Significant departures from this approximation are considered in Section \ref{sec: scaling}.

\subsection{Analytical merons}

To reinforce the results from the previous section, we perform an analytical calculation of a `model' meron (or anti-meron) in Co, stabilised by the presence of a planar vortex (or anti-vortex) in the adjacent AFM oxide, assuming very simple functional forms for the $z$ component of the magnetisation.  We demonstrate that the characteristic size of the meron `core' is indeed proportional to $L_{ex.b}$.  In this calculation, we disregard the effect of magneto-static energy (included in our micromagnetic model --- see below), so the calculation is exactly identical for a vortex$\vert$meron and anti-vortex$\vert$anti-meron.  

The general expression for the normalised meron magnetisation is

\begin{eqnarray}
\label{eq: meron_spins}
m_x&=&-\sin \psi \sin \phi\nonumber\\
m_y&=&\sin \psi \cos \phi\nonumber\\
m_z&=&\cos \psi
\end{eqnarray}

where $\phi$ is the polar angle and

\begin{eqnarray}
\label{eq: meron_definition}
\tan \psi&=&\mathcal{F}\left(\frac{r}{R}\right)\nonumber\\
\cos \psi&=& \frac{1}{\sqrt{\mathcal{F}^2+1}}\nonumber\\
\sin \psi&=& \frac{\mathcal{F}}{\sqrt{\mathcal{F}^2+1}}
\end{eqnarray}

Here, $\mathcal{F}(x)$ is a continuous function with $\lim_{x \rightarrow 0} \mathcal{F}(x)=0$ and $\lim_{x \rightarrow \infty} \mathcal{F}(x)=\infty$, and $R$ is a characteristic scale.  In Appendix \ref{sec: meron_calculations} we provide calculations for a number of simple cases, including the `projective' meron ($\mathcal{F}(x)=x$), which can be obtained by projecting a `hairy' sphere of radius $R$ from its centre onto a tangent plane \cite{Rajaraman1987}, and the more general case in which $\mathcal{F}(x)$ is a polynomial.  In order to provide a direct comparison with the linear domain wall, we also discuss the case of the `linear meron', where the magnetic moment is entirely in plane outside a radius $R$, while inside this radius it rotates uniformly towards the centre of the meron, where it is aligned along $z$.  As shown in Appendix \ref{sec: meron_calculations}, the `projective' meron case is unstable, due to the logarithmic energy cost owing to the swirling spins at large distances, while in all other cases the width of the meron core scales with the exchange bias length $L_{ex.b}$ (see also eq. \ref{eq: meron core width}):

\begin{equation}
\label{eq: meron core width1}
W_{core}=2 \kappa \sqrt{ \frac{A_{ex}d}{K_{ex.b} }}=2 \kappa L_{ex.b}
\end{equation}  

with $\kappa \approx 1$.

\section{Micro-magnetic modelling}
\label{section: Micro-magnetic modelling}

\subsection{Micromagnetic modelling in OOMMF}

\begin{figure}
\centering
\includegraphics[scale=1.5, trim={0cm 0cm 0 0},clip]{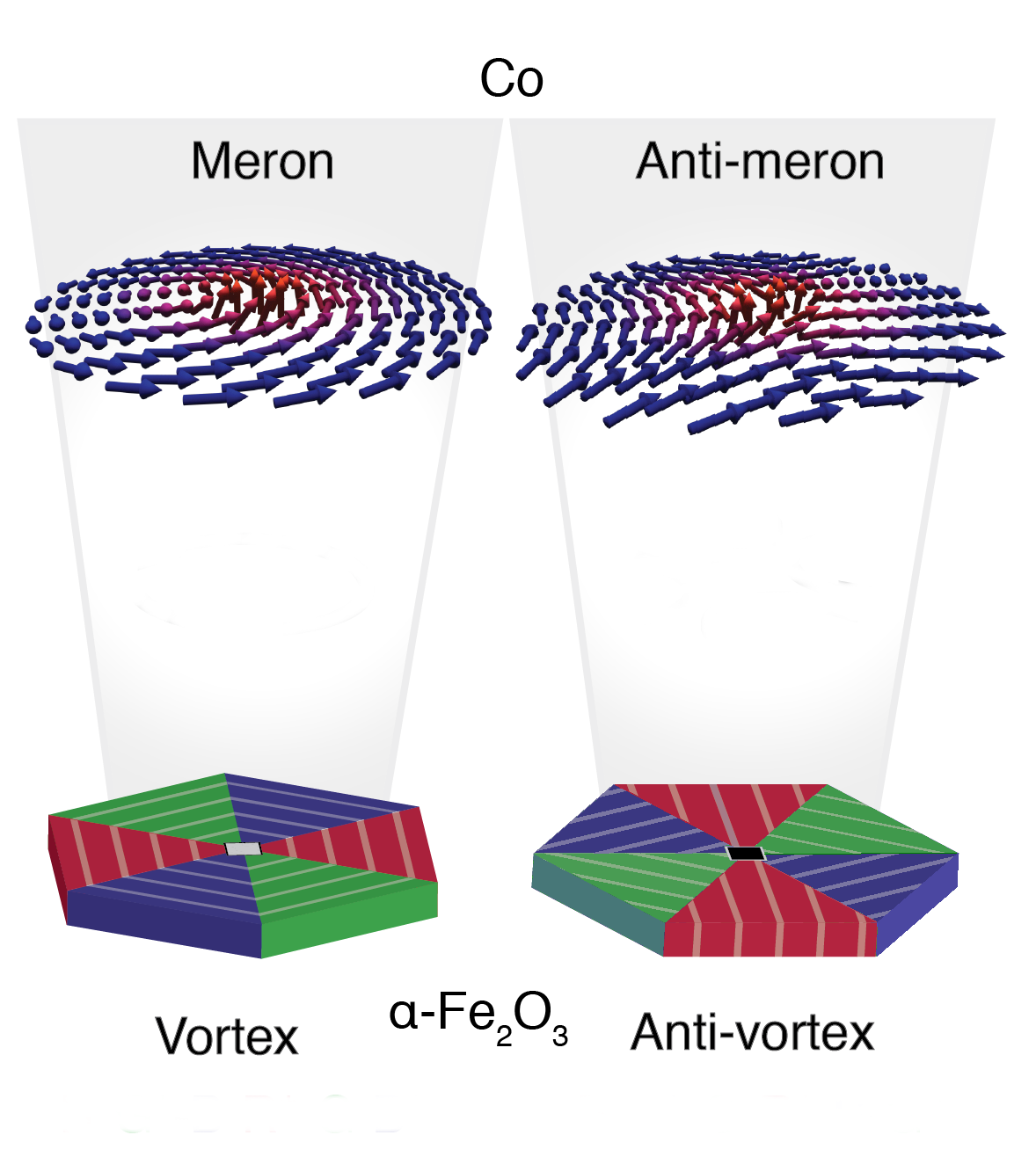}
%left lower right upper
\caption{(Colour online): Overview of the coupled \FeO\ $\vert$ Co system that was simulated in this section.  The bottom layer is \FeO\, and the figure shows a vortex (bottom left) and an anti-vortex (bottom right).  Light-coloured lines indicating the direction of the staggered magnetisation.  Only the top termination of \FeO\ was included as a fixed layer in the simulations (Figure \ref{fig: wedges}, left panel).  Spins in the top Co layer were set to a random orientation prior to the start of the simulation, and develop meron/anti-meron structures at the end of the simulation, as shown in the top panels.}  
\label{fig: system_description}
\end{figure}

\begin{figure}
\centering
\includegraphics[scale=0.45, trim={0cm 0cm 0 0},clip]{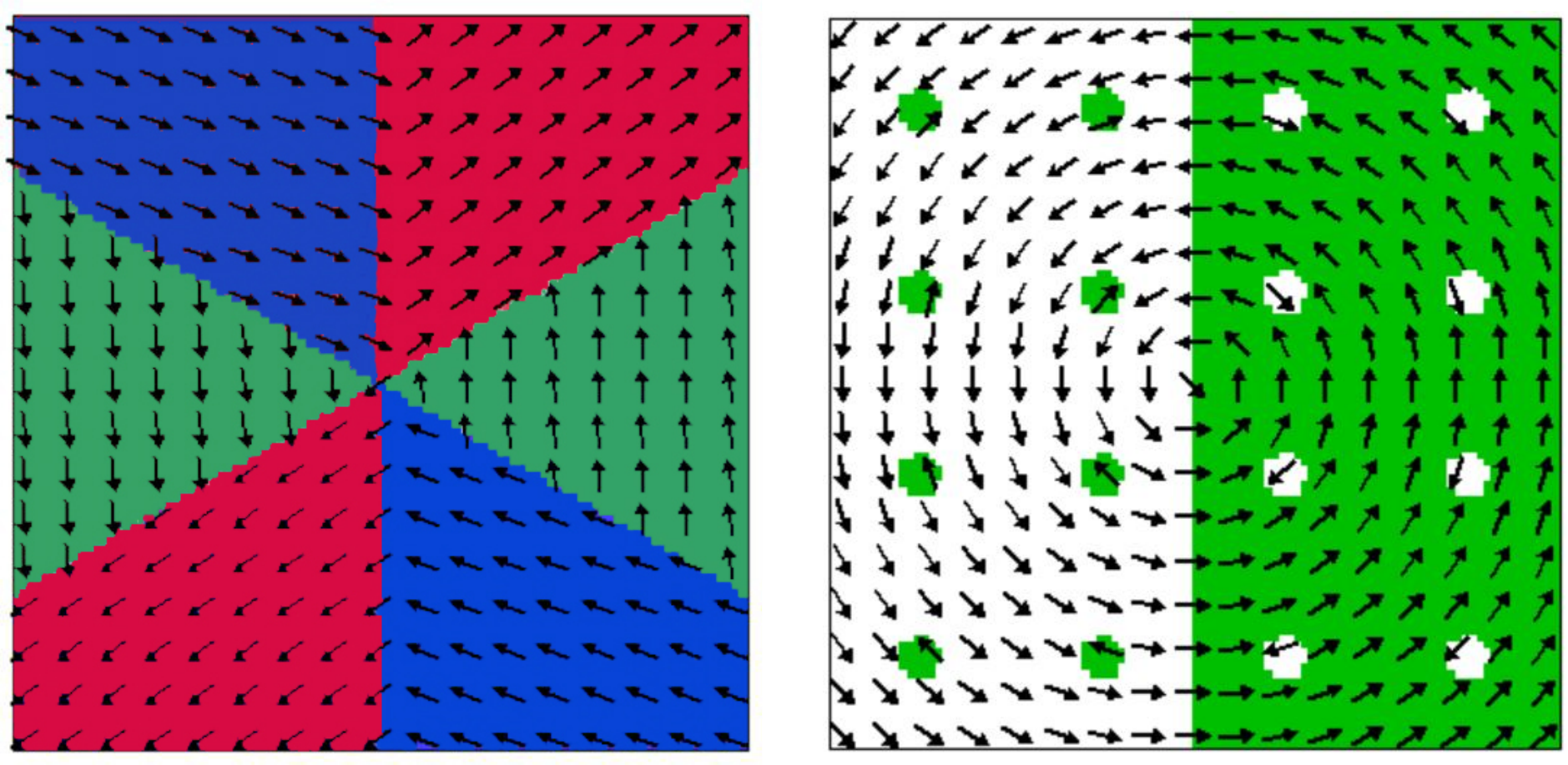}
%left lower right upper
\caption{(Colour online): Termination layers of AFM \FeO: \textbf{Left}: Antivortex in \FeO\ with six distinct wedges (colours/grayscale), as observed in \cite{Chmiel2018}.  The simulated area is 200 $\times$ 200 \si{\nano\meter\squared}. \textbf{Right} Vortex with uniformly rotating magnetisation and circular regions of magnetisation reversal, introduced to simulate surface roughness.  The diameter of the circular regions  (emphasised by the shading) is 6 \si{\nano\meter}, while the simulated area is 100 $\times$ 100 \si{\nano\meter\squared}.  Colours/shading emphasise the moment directions.}  
\label{fig: wedges}
\end{figure}

Micromagnetic simulations were performed using the program OOMMF \cite{Donahue1999}.  No periodic boundary conditions were employed and the AFM layers were kept fixed throughout the simulations in all cases.   A general overview of the system we simulated is provided in Figure \ref{fig: system_description}.  Uniform \FeO\ termination layers were described as having magnetisation of constant magnitude, which rotates counterclockwise (for vortices) or clockwise (for anti-vortices) when moving on a counterclockwise path around the centre.  The total simulated area was 200 $\times$ 200 \si{\nano\meter\squared}  and the discretisation cell sizes were $D_{xy}$ = 2 \si{\nano\meter} and $D_z$ =  1 \si{\nano\meter}, respectively, with the \FeO\ layer being 1 cell thick.  We performed simulations both with uniformly rotating magnetisation and also with constant magnetisation within six equal wedges, which reproduce the experimental images of AFM vortices/anti-vortices \cite{Chmiel2018} (Fig. \ref{fig: wedges} \textbf{left}).   The meron structures in the two cases are extremely similar, although the sharp AFM boundaries associated with the wedges introduce N\'eel domain walls (see below).  To model the effect of surface roughness, an additional set of simulations were performed over 100 $\times$ 100 \si{\nano\meter\squared} with $D_{xy}$ = 0.5 \si{\nano\meter} in-plane discretisation, with the \FeO\ magnetisation being reversed within circular islands arranged on a regular grid (Figure \ref{fig: wedges}\textbf{ right}).

Co layers of different thickness (1 -- 8 \si{\nano\meter}) were placed in direct contact with the \FeO\ layer and interacting with it through an exchange stiffness $A^{\rm{Fe-Co}}_{ex}$ (which is not known \emph{a priori} --- see below and Appendix \ref{app: exchange_bias} for a full discussion), so a series of simulations were performed spanning a wide range of $L_{ex.b}$.  For $A_{ex}^{\rm{Co-Co}}$, we have used the literature value of 18 \si{\pico\joule\per\meter} \cite{Abo2013}.  The Co magnetisation is assumed to be 1.4 $\times$ 10$^6$ \si{\joule\per\tesla}, yielding a magnetostatic exchange length $L_{ms}=\sqrt{2A_{ex}/\mu_0M^2}=3.8$ \si{\nano\meter} \cite{Abo2013}.  The magnetisation in each cell was initially set at a random orientation, and it was then allowed to evolve according to the Landau-Lifshitz-Gilbert equation \cite{Gilbert2004} until a stable configuration was attained.  Since here we are not interested in magnetisation dynamics, the dimensionless damping factor $\alpha$ should not influence the outcome; in our simulations, $\alpha$ was set to 0.5 --- a value that was empirically found to yield good convergence properties of the model. The simulation time step was adjusted by the programme in the range 1-100 \si{\pico\second},  while the convergence criterion was 5\si{\degree\per\nano\second}. Unless $A^{\rm{Fe-Co}}_{ex}$ was set to a very small value, the Co magnetisation always formed a meron/anti-meron, with the same vorticity as the underlying vortex/antivortex in \FeO\, while the core magnetisation was randomly up or down in each simulation run.

Although strictly a technical issue, the implementation of exchange bias in our micromagnetic simulations deserves a separate remark, since in OOMMF it is not possible to introduce the equivalent of $K_{ex.b}$ directly.  Instead, the effect of exchange bias can be reproduced by employing an exchange stiffness parameter $A^{\rm{Fe-Co}}_{ex}$, which acts only on the interface cells between \FeO\ and Co.  The only \emph{caveat} is that $A^{\rm{Fe-Co}}_{ex}$ is not a physical parameter, since it depends on the size of the discretisation cell $D_z$ along the $z$ direction, as discussed at length in Appendix \ref{app: exchange_bias}.  The scaling $K_{ex.b}=2 A^{\rm{Fe-Co}}_{ex} /D_z$ between $A^{\rm{Fe-Co}}_{ex}$ and the physical parameter $K_{ex.b}$ , derived in Appendix \ref{app: exchange_bias}, was verified in series of simulations with different $D_z$.

\subsection{Modelling MFM and N-V centre microscopy images}

Simulated MFM and diamond N-V centre microscopy images were produced from meron/anti-meron structures obtained by setting $L_{ex.b}=L_{ms}$ as a representative value.

For MFM, we employed the phase shift method, whereby the image is generated based on the shift in phase between the drive and the cantilever, which is driven close to resonance \cite{Hartmann1999}.  The phase shift is given by the formula:

\begin{equation}
\label{eq: phase_shift}
\Delta \Phi=-\frac{Q}{k}\frac{\partial F_z}{\partial z}
\end{equation}

where 

\begin{equation}
\vec{F}=\nabla(\bm{\mu} \cdot \vec{H})
\end{equation}

is the force on the cantilever tip due to the stray magnetic field $\vec{H}$, and $\bm{\mu}$ is the magnetic moment of the tip. The derivative of the force was calculated numerically and averaged over a number of `voxels' comprising the shape of the tip. The magnetic moment of the tip was kept constant at $|\bm{\mu}|$ = 1.2 $\times$ 10$^{-19}$ \si{\joule\per\tesla}, whilst different tip sizes and shapes were tested.  The cantilever spring constant in equation \ref{eq: phase_shift} was $k=$ 2.8 \si{\newton\per\meter}, while the quality factor $Q$ was set at 100, which is much less than the `bare' cantilever $Q$ but is realistic for room-temperature measurements in the presence of a water film. 

For diamond N-V centre microscopy, a first set of images were produced without bias magnetic field, assuming that the signal is proportional to the magnitude of the projection of the stray magnetic field along the direction of the defect, which was aligned with the [111] crystallographic direction of the diamond \cite{Rondin2014, Gross2017}.  The [001] and [110] crystallographic directions of the diamond were aligned along the $z$ and $x$ axes, respectively.  A second set of images was produced with a bias field of $\sim$ 110 \si{\milli\tesla} along the $x$ direction, such that the projection of the stray plus bias magnetic field along the defect never changes sign.  This field should be consider an upper limit of what it is possible to apply experimentally, since a field of this magnitude on the surface of the sample is likely to cause meron annihilation \cite{Chmiel2018}.

\section{Results}
\label{sec: Results}

\subsection{Meron/anti-meron formation and features size}

\begin{figure}
\centering
\includegraphics[scale=0.40, trim={0cm 0cm 0 0},clip]{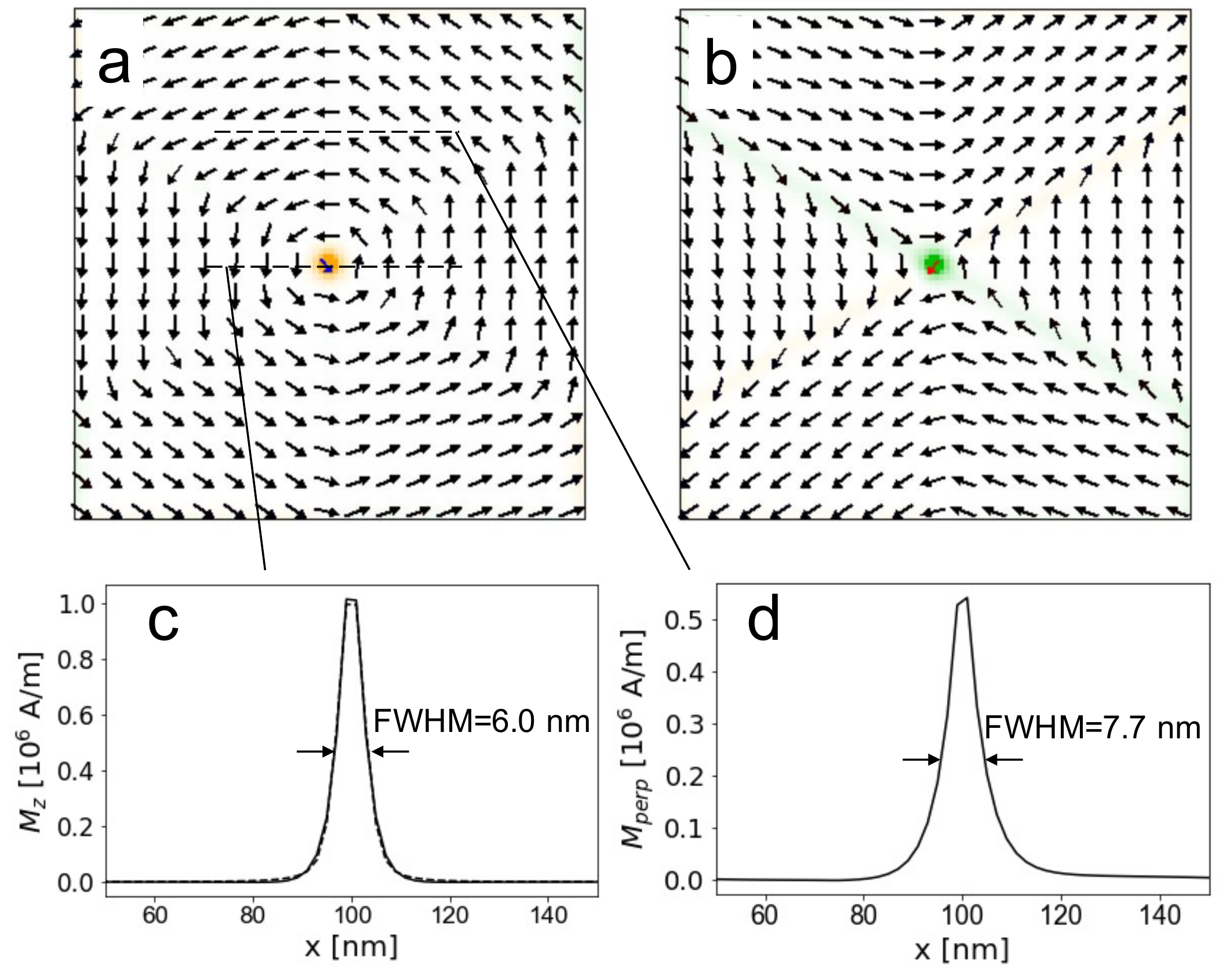}
%left lower right upper
\caption{(Colour online): \textbf{Top panels}: 200 $\times$ 200 \si{\nano\meter\squared} OOMMF simulations of a meron (\textbf{a}) and anti-meron (\textbf{b}) in a 1 \si{\nano\meter} Co film, with an `exchange bias' field from a hexagonal vortex/antivortex in \FeO.  The exchange bias length for the simulations was set at $L_{ex.b}$ = 3.8 \si{\nano\meter}.  Away from the (anti-)meron core, Co spins and Fe spins are parallel.  The in-plane spin component is indicated by the arrows, while the out-of-plane ($z$) components is  in shaded colour (grayscale). \textbf{Bottom panels}: profiles of the magnitudes of Co spin component along $z$ through the meron core (\textbf{c}) and of Co spin component orthogonal to the Fe spins through a N\'eel domain boundary (\textbf{d}).} 
\label{fig: vortex/antivortex}
\end{figure}

Figure \ref{fig: vortex/antivortex} shows a typical `converged' Co spin configurations for a meron (\textbf{a}) and an anti-meron (\textbf{b}) stabilised by an AFM vortex/anti-vortex, similar to that in Figure \ref{fig: wedges} (\textbf{left}), using exchange lengths $L_{ex.b}$ =  $L_{ms}$ = Ô$L_{ex.b}$ = 3.8 \si{\nano\meter}.  Figure \ref{fig: vortex/antivortex}\textbf{c} shows the $z$ component of the magnetisation plotted along a line cutting through the meron core, while figure \ref{fig: vortex/antivortex}\textbf{d} shows the component of the magnetisation \emph{perpendicular} to the underlying AFM spins, plotted along a line cutting through a N\'eel domain wall (lines shown in Figure \ref{fig: vortex/antivortex}\textbf{a}).\footnote{This is a natural generalisation of the usual 180\si{\degree} domain wall model to the case of a trigonal system having 60\si{\degree} domains. It follows that, for a constant-magnetisation 60\si{\degree} domain wall, the perpendicular component is at most 1/2 of the total magnetisation.}  At the centre of the meron/antimeron, the magnetisation is completely aligned along the $z$ axis.  For the meron,  $M_z$ is non-zero only near the core, while for the anti-meron there is a sizeable  $M_z$ component along the two diagonal lines where the in-plane magnetisation is along the radial direction.  Interestingly, the full width at half maximum (FWHM) of the meron core (6 \si{\nano\meter}) is smaller that that of the N\'eel domain wall (7.7 \si{\nano\meter}) (this discrepancy is qualitatively consistent with the analytical results in Appendix \ref{app: linear_antiphase} and \ref{sec: meron_calculations}). 

\subsection{Meron/anti-meron core scaling}
\label{sec: scaling}

Having established the basic procedure to produce micromagnetic simulations of merons and anti-merons, we proceeded to generate a series of structures with different values of $L_{ex.b}$, whilst keeping $L_{ms}$ at the literature value of 3.8 \si{\nano\meter}.  In establishing an appropriate range for $L_{ex.b}$, one should consider that, for an ideal system, the exchange stiffness and interface energies are related to the microscopic exchange constants $J$ by the following equations (see ref. \cite{Schollwock2004}):

\begin{eqnarray}
A^{\rm{Co-Co}}_{ex}&=&c_1\frac{J^{\rm{Co-Co}} S_{\rm{Co}}^2}{a_{nn}}\nonumber\\
K^{Co-Fe}_{ex.b}&=&c_2\frac{J^{Co-Fe} S_{\rm{Co}}S_{\rm{Fe}}}{a_{nn}^2}
\end{eqnarray}

where $a_{nn}$ is the atomic nearest-neighbour distance, $S_{\rm{Fe}}$ and $S_{\rm{Co}}$ are the cobalt and iron spins, while $c_1$ and $c_2$ are small numbers that depend on the coordination and $c_1 > c_2$.  In the approximation of equal exchange constants and spins, for an ideal system one would have

\begin{equation}
L_{ex.b}\approx \sqrt{\frac{c_1}{c_2}d \,a_{nn}}
\end{equation} 

so, for $d$ = 1 \si{\nano\meter} it is reasonable to take 1-2 \si{\nano\meter} as the \emph{lower bound} for $L_{ex.b}$.  In a real system,  one would expect that $K^{Co-Fe}$ should be significantly weakened by surface roughness, so we tested much larger values of $L_{ex.b}$ (up to  35 \si{\nano\meter}), up until the point where merons/anti-merons ceased to be stable.  

\begin{figure}
\centering
\includegraphics[scale=0.34, trim={0cm 0cm 0cm 0},clip]{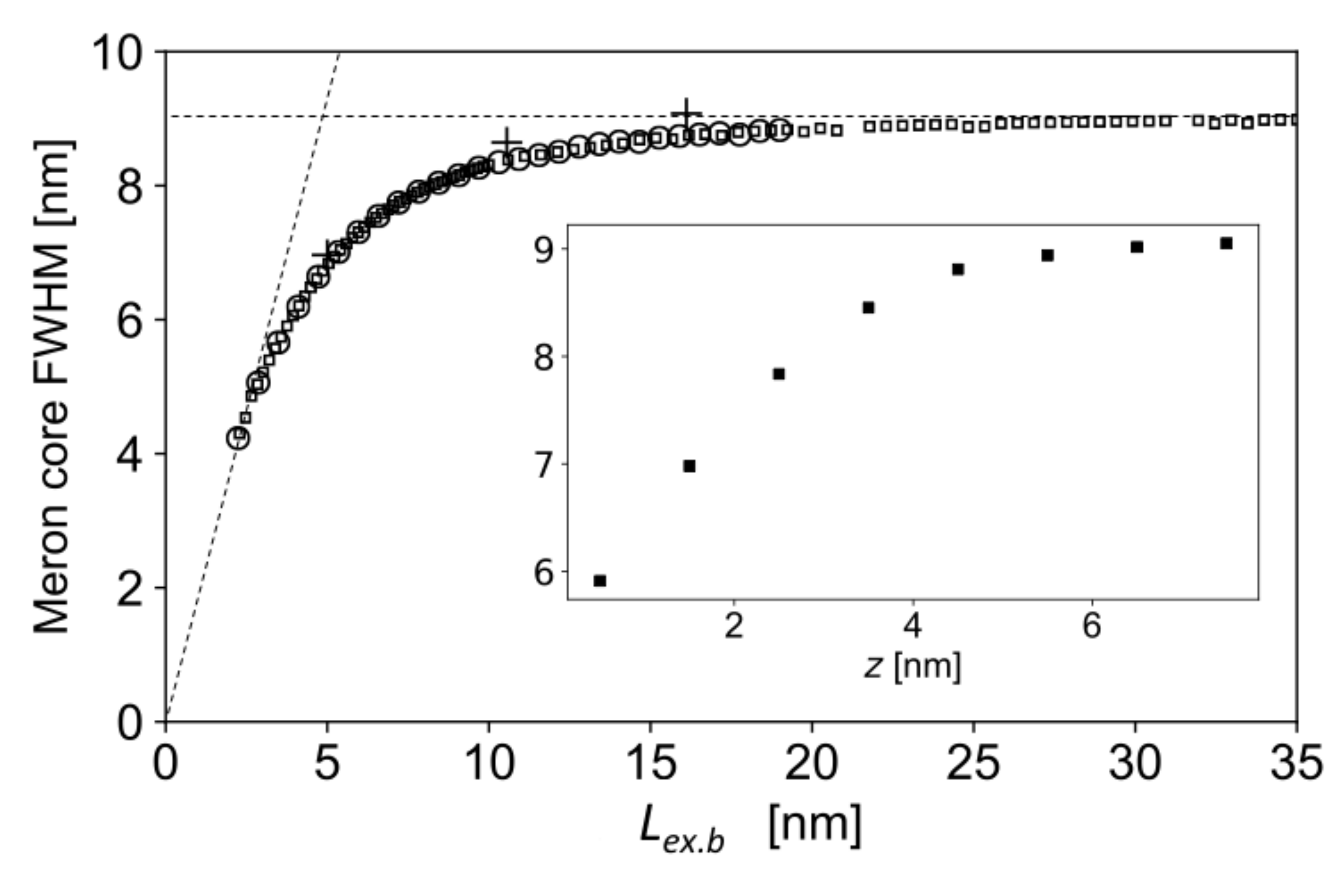}
\caption{\textbf{Main panel}.  Full width at half maximum of the meron core as a function of the exchange-bias length $L_{ex.b}$, as defined in eq. \ref{eq: ex_l_def}.  The anisotropy length $L_{ms}$ was set at 3.8 \si{\nano\meter} for all data points.  \textit{Open squares} and \textit{open circles} correspond to a Co film thickness $d$ of 1 \si{\nano\meter} and 2 \si{\nano\meter}, respectively, while crosses are for anti-merons with $d = $1 \si{\nano\meter}.  The dashed lines corresponds to $\sim$ 2.37 $L_{ms}$ (horizontal) and 1.67 $L_{ex.b}$ (diagonal).  \textbf{Inset}.  Meron core FWHM versus distance $z$ from the interface for a 8 \si{\nano\meter} Co film with $L_{ex.b}=3.8$ \si{\nano\meter}.} 
\label{fig: scaling}
\end{figure}

Once the models had converged, the meron/anti-meron cores were fitted by 2-dimensional pseudo-Voigt functions, which enabled the FWHM to be extracted systematically.  The results of these fits are summarised in figure \ref{fig: scaling} (\textbf{main panel}).  For very strong exchange bias interactions (small values of $L_{ex.b}$), interface physics is dominant, and the core size is proportional to  $L_{ex.b}$, consistent with our analytical calculations (Appendix \ref{sec: meron_calculations}).  In fact, the proportionality constant extracted from the initial slope of the plot ($\approx$ 1.67) is rather close to the analytical value of 1.5 (equation \ref{eq: fwhm_core}).  For larger values of $L_{ex.b}$, the core size in increasingly dominated by `bulk' physics, and eventually saturates at $\approx $ 2.37 $L_{ms}$.  Meron and anti-meron core sizes are almost identical for small $L_{ex.b}$, as one would expect, but anti-meron cores are slightly bigger for larger $L_{ex.b}$, consistent with the fact that anti-vortices have very unfavourable magneto-static energies.   Compared to merons, anti-meron ultimately become unstable for smaller values of $L_{ex.b}$.

For small film thicknesses (1-2 \si{\nano\meter}), the magnetisation is essentially independent of $z$ and the effect of the thickness $d$ can be included in the definition of $L_{ex.b}$ given by equation \ref{eq: ex_l_def}. For thicker films, this ceases to be true, as shown in the inset of figure \ref{fig: scaling}, which demonstrates the transition from `surface' to `bulk' physics \emph{within the same film}.  In the example shown ($L_{ex.b}=3.8$ \si{\nano\meter}), the meron core is compact in the portion of the film closer to \FeO\, but `flares out' as $z$ increases, until it saturates to the `bulk' value of $\sim $ 2.37 $L_{ms}$.

One conclusion of this section is that the meron/anti-meron core size in Co \emph{never exceeds} $\sim$ 9 \si{\nano\meter} \emph{regardless} of the strength of the interface interaction and the film thickness.  This has very important implications for the possibility of creating dense meron/anti-meron networks (see discussion at the end of the paper).  A second observation is that, based on our simulations, there is likely to be a difference in the pinning strength required to keep merons and anti/merons pinned to \FeO, due to their different magnetostatic energy.  This feature is amenable to be exploited for applications, for example, to `unpin' one type of particle selectively.

\subsection{Modelling surface roughness}

\begin{figure}[h]
\centering
\includegraphics[scale=0.7, trim={0cm 0cm 0cm 0},clip]{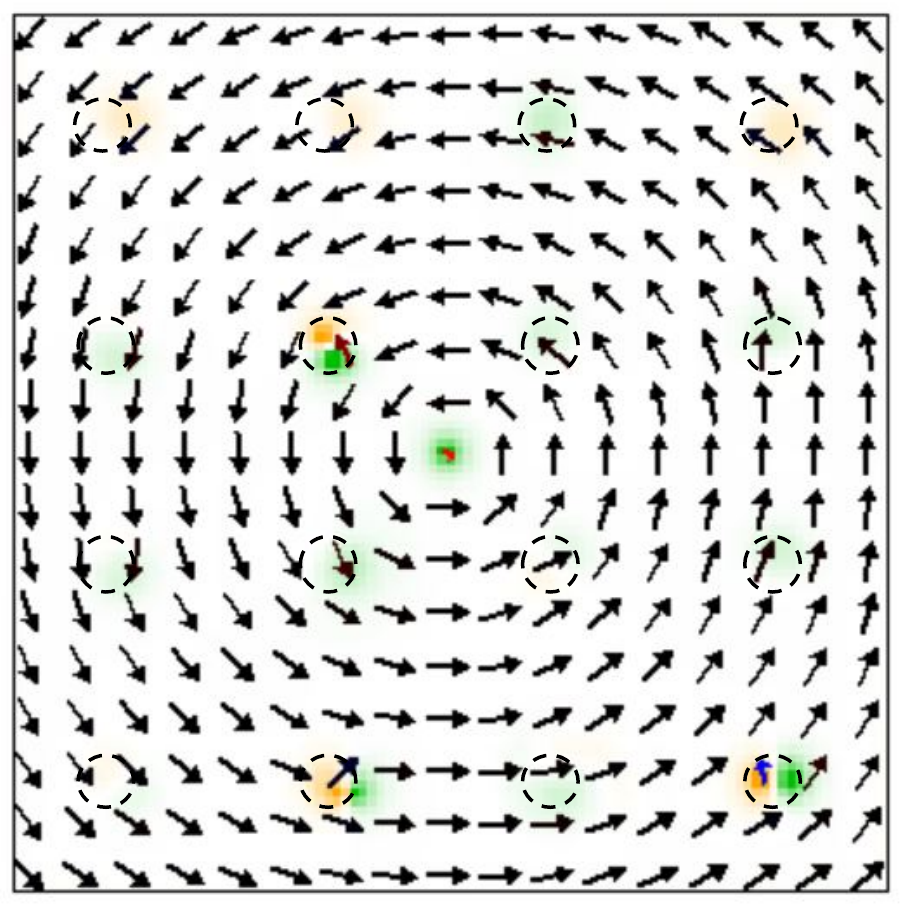}
\caption{Effect of \FeO\ roughness on the meron structure.  200 $\times$ 200 \si{\nano\meter\squared} OOMMF simulation of a meron  in a 1- \si{\nano\meter} - thick Co layer, stabilised by a `rough' \FeO\  interface.  Roughness is simulated by introducing disk-like areas of AFM spin reversal, with a diameter of 6 \si{\nano\meter} (dotted lines). The exchange lengths were set at $L_{ex.b}=L_{m.s}=3.8$ \si{\nano\meter}.  Color intensity is proportional to the out-of-plane component of the Co moments.} 
\label{fig: roughness}
\end{figure}

As previously mentioned, (see section \ref{sec: scaling}), our initial assumption of a uniform FM termination for \FeO\ cannot be realistic, since in all but the most perfect epitaxial films there is always a degree of surface roughness.  One may even question whether merons/anti-merons can be stabilised in the presence of a rough \FeO\ interface, since the sign of the magnetisation in the layer in direct contact with Co changes sign in different termination layers.  Intuitively, one would expect the lateral scale of the termination terraces to be an important parameter:  features in Co cannot be smaller than 1-2 times the relevant exchange length, so the effect of fine-grained roughness should be to weaken the dominant exchange-bias interaction without altering the topology of the Co features.  This intuition is confirmed by our micromagnetic models (shown in figure \ref{fig: roughness}), in which surface roughness is simulated by regions of \FeO\ spin inversion in the shape of circular `terraces' of 6 \si{\nano\meter} diameter.  In order to prevent the roughness-related features from being `washed out' by finite-scale effects, these simulations were performed on smaller discretisation cells (0.5 \si{\nano\meter}).  As evident from figure \ref{fig: roughness}, the shape of the meron structure in Co is largely unaffected by our model roughness.  The main effect of the terraces is to introduce a small local distortion and a non-zero $z$ component of the magnetisation --- a very reasonable result, since this lowers the exchange bias energy at the terrace site.

\subsection{MFM imaging and N-V centre imaging}

\begin{figure}[h]
\centering
\includegraphics[scale=0.45, trim={0.2cm 0cm 0cm 0},clip]{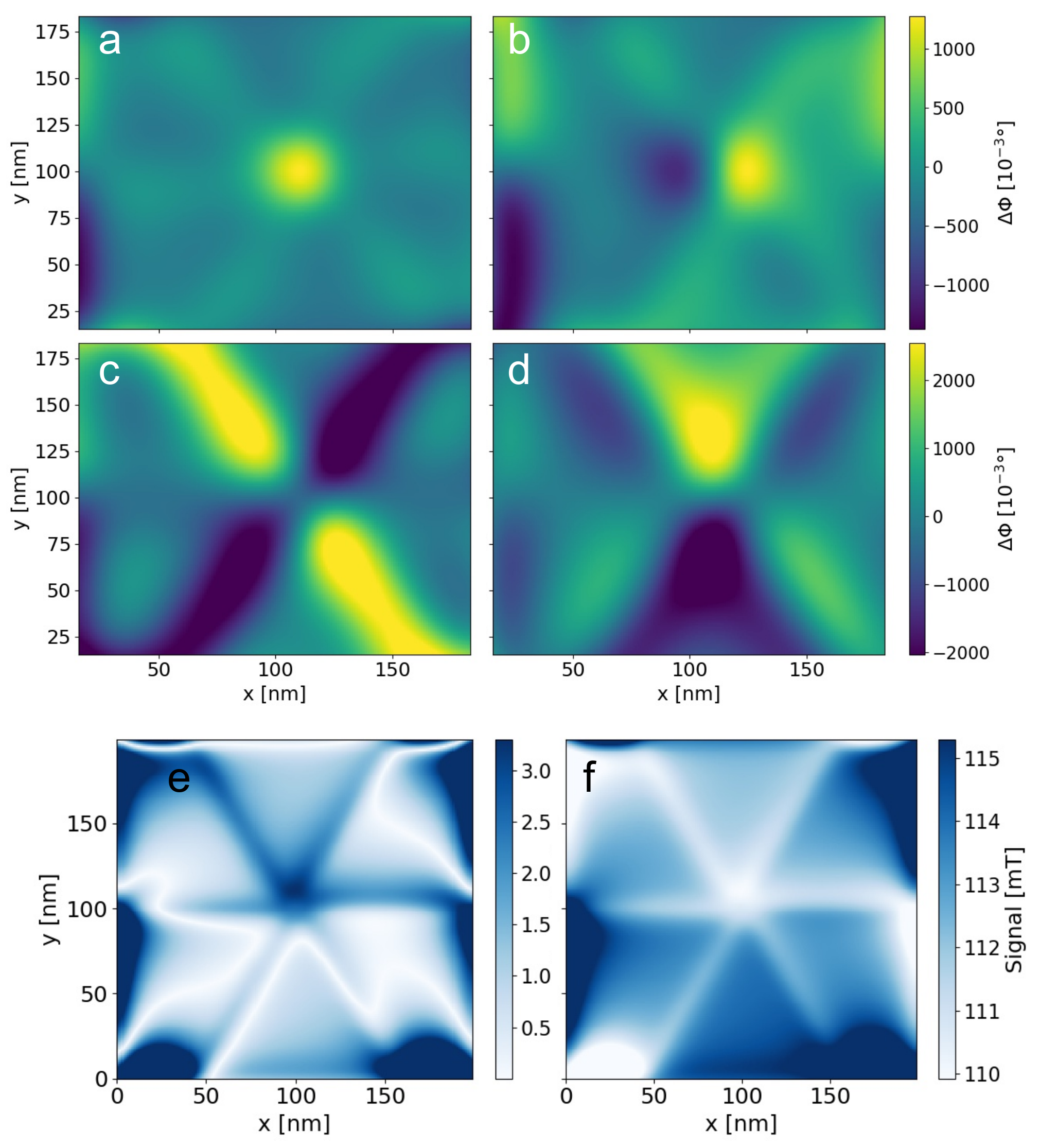}
\caption{\textbf{Top and middle row}: Simulated MFM images of merons (top) and anti-merons (middle) with both $L_{ex.b}$ and $L_{ani}$ parameters set to 3.8 \si{\nano\meter}. In all cases, the magnetic moment of the tip was $1.15\times10^{-22}$ \si{\joule\per\tesla} and the working distance was 20 \si{\nano\meter}.  \textbf{a} and \textbf{c}: tip magnetised out of the page;  \textbf{b} and \textbf{d}: tip magnetised in the y direction.  \textbf{Bottom row}: images produced using an N-V centre tip with the x axis of the sample aligned parallel to the in-plane projection of the defect, at a working distance of 11 \si{\nano\meter}; \textbf{e}: no bias magnetic field; \textbf{f}: bias field of $\sim$ 110 \si{\milli\tesla} along the $x$ direction.  } 
\label{fig: mfm_nv}
\end{figure}

Figure \ref{fig: mfm_nv} \textbf{a}-\textbf{d} show simulated MFM images of a meron (\textbf{a},\textbf{b}) and an anti-meron (\textbf{c},\textbf{d}), at a tip-to-film working distance of 20 \si{\nano\meter}, which is typical for this technique. The tip was modelled as a pyramid with dimensions 31 $ \times$ 31 \si{\nano\meter\squared} base $ \times$ 31 \si{\nano\meter} height, and a total magnetic moment of 1.2 $\times$ 10$^{-19}$ \si{\joule\per\tesla}.  Images were produced with both perpendicular (\textbf{a},\textbf{c}) and in-plane (\textbf{b},\textbf{d}) magnetisation of the tip.  The meron core is distinguishable within typical instrumental sensitivity, albeit significantly broadened by resolution effects.  With the tip magnetisation perpendicular to the film (figure \ref{fig: mfm_nv} \textbf{a}), the core appears as a disk-shaped area of phase shift, and could be confused with other MFM features of different origin.  By contrast, when the tip is magnetised in plane, the core displays a typical region of phase inversion (\ref{fig: mfm_nv} \textbf{b}), which could be used as a characteristic signature.  Somewhat surprisingly, for anti-merons (figure \ref{fig: mfm_nv} \textbf{c},\textbf{d}), the $X$-shaped ridge structure in the stray field is a much more prominent and recognisable feature than the core for both perpendicular and parallel tip magnetisation.  

Figure \ref{fig: mfm_nv} \textbf{e}-\textbf{f} show simulated N-V centre microscopy images of a meron, taken without (\textbf{e}) and with (\textbf{f}) a bias field in the direction of the defect axis.  The working distance between the surface and the N-V centre was 11 \si{\nano\meter}, which is realistic for a shallow defect.  Because this technique is directly sensitive to the amplitude of the stray field, edge effects representing an artefact of the 200 $\times$ 200 \si{\nano\meter} simulation region are very prominent in the simulated images.  Nevertheless, details of the meron structure are very evident and are much less broadened by resolution effects than for MFM.  In addition to the tight meron core, one can also clearly distinguish the N\'eel domain walls, which were all but invisible in MFM.  Both unbiassed and field-biassed images are useful and provide complementary information, which can help unravel the magnetic structure of the meron.  The N-V centre microscopy technique seems therefore very promising as an alternative and complement to X-PEEM, which has thus far been used exclusively to image these structures.

\section{Conclusions}

In conclusion, we have produced micromagnetic models and simulated MFM/N-V microscopy images of coupled (anti-)vortex/(anti-)meron structures in \FeO$\vert$Co heterostructures. Perhaps the most important conclusion of our analysis is that meron/anti-meron cores in Co remain small ($<$ 10 \si{\nano\meter}) even when the exchange-bias interaction between AFM and FM layers is extremely weak.  The fundamental reason for this is that the crossover between `surface' and `bulk' phenomenology (at strong and weak exchange-bias interactions, respectively) is controlled by two different length-scales, and that the bulk-related magnetostatic length keeps the FM features small even when the surface-related exchange-bias length is long.  One outcome of this is that \FeO$\vert$Co heterostructures and similar systems could, in principle, support very dense topological networks even in the presence of rough interfaces, which tend to weaken the net exchange bias interaction.  This is of course precisely what is wanted for applications, for example, in high-density magnetic storage.

One obstacle to fast-track development of these systems is the requirement for scarce X-PEEM beamtime at synchrotron sources to characterise the AFM and FM topological structures.   Our simulated MFM and N-V microscopy images demonstrate the existence of characteristic features associated with FM merons and anti-merons, which could be used to complement X-PEEM with much more accessible, lab-base techniques.

\section{Acknowledgements}

Work done at the University of Oxford is funded by EPSRC Grant No. EP/M020517/1, entitled Oxford Quantum Materials Platform Grant. R. D. J. acknowledges support from a Royal Society University Research Fellowship.  We thank S. Parameswaran for discussions and Tom Lancaster (University of Durham) and Hariom K. Jani (National University of Singapore) for commenting on the manuscript. 

\appendix

\section{Exchange-bias parameter and micromagnetic scaling}
\label{app: exchange_bias}

In the exchange-bias calculations described in section \ref{sec: features size}, we have employed the parameter $K_{ex.b}$ together with the definition of the exchange bias energy:

\begin{equation}
\label{eq: energy_per_area1}
E_{ex.b}=\int d\sigma \, K_{ex.b} (1- \cos \theta)
\end{equation}

where $\theta$ is the angle between the AFM and the FM spins at the interface.  Although in the OOMMF micromagnetic implementation it is not possible to introduce the equivalent of $K_{ex.b}$ directly, its effect can be reproduced by employing an exchange stiffness parameter $A^{\rm{Fe-Co}}_{ex}$, which acts only on the interface cells between \FeO\ and Co.  In order to obtain a correct scaling of the model, one must be able to relate $A^{\rm{Fe-Co}}_{ex}$ (which, as we shall see, is scale-dependent) with the `physical' parameter  $K_{ex.b}$.

If at the interface the angle between the spins in \FeO\ and those in Co is $\theta$, the discrete gradient term is:

\begin{equation}
\left(\nabla \right)^2=\frac{1}{D_z^2}\left(\left(\cos \theta-1\right)^2+\left(\sin \theta\right)^2\right)=\frac{2}{D_z^2} \left (1-\cos \theta\right)
\end{equation}

where $D_z$ is the length of the discretisation cell along the $z$ axis.  The energy per unit area is therefore

\begin{eqnarray}
\frac{\partial E_{ex.b}}{\partial \sigma}&=&2 A^{\rm{Fe-Co}}_{ex} \frac{2}{D_z^2} \left (1-\cos \theta\right) \frac{D_zD_{xy}^2}{D_{xy}^2}\nonumber\\
&=&2 \frac{A^{\rm{Fe-Co}}_{ex} }{D_z} \left (1-\cos \theta\right) 
\end{eqnarray}

where $D_{xy}$ is the length of the discretisation cell in the plane of the film.  This is identical to the expression in eq. \ref{eq: energy_per_area1} (see also eq. \ref{eq: energy_per_area}) with the identification $K_{ex.b}=2 A^{\rm{Fe-Co}}_{ex} /D_z$.  By performing simulations with different discretisation cell sizes, we have verified that this is indeed the correct scaling factor to be applied for obtaining the same feature sizes in simulations with different $D_z$.

Expressions such as eq. \ref{eq: domain wall width} and \ref{eq: meron core width1} would also enable a value for $K_{ex.b}$ to be estimated from the feature sizes of experimental images, assuming that they are not limited by instrumental resolution.  

\section{Exchange-bias domain walls}
\label{app: linear_antiphase}

\begin{figure}[h]
\centering
\includegraphics[scale=0.4, trim={0cm 0cm 0cm 0},clip]{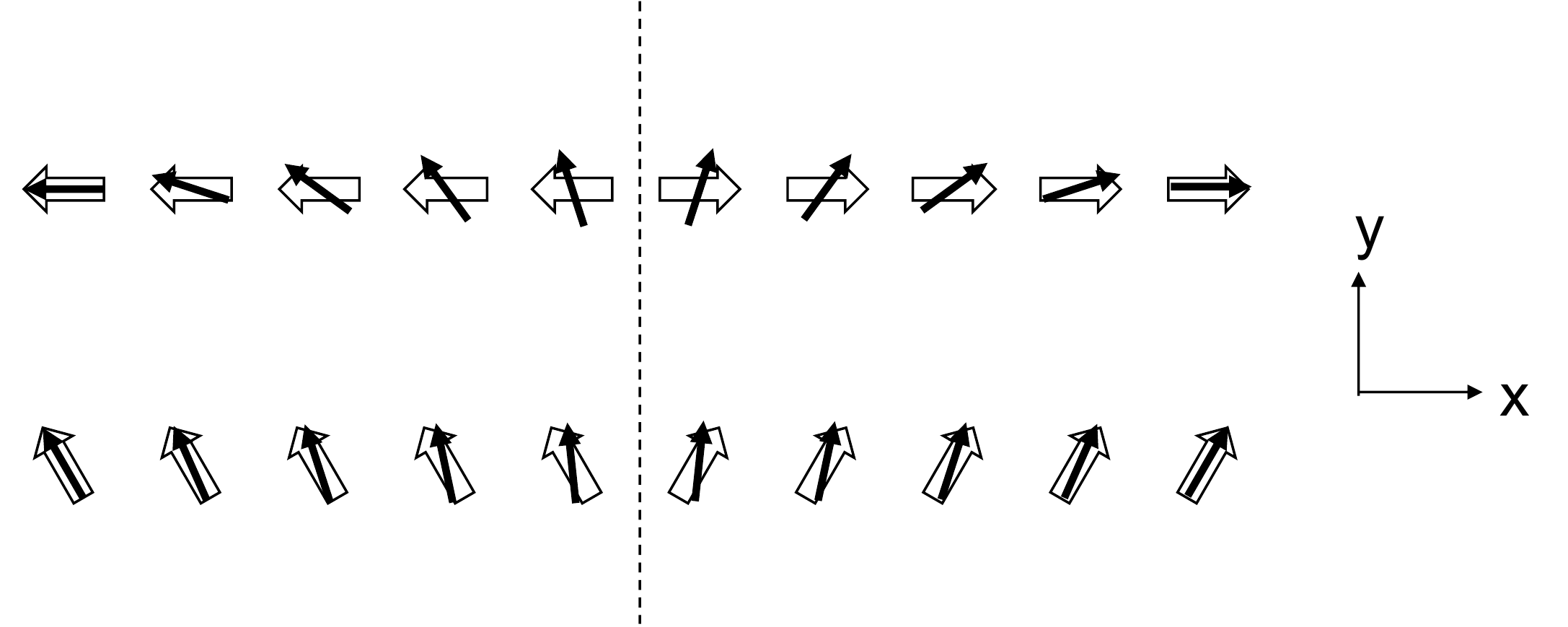}
\caption{Linear domain boundaries in Co (black arrows) induced by a sharp domain boundary in \FeO\ (white arrows representing the top uncompensated layer). \textbf{top}: 180\si{\degree} domain boundary.  \textbf{bottom}: 60\si{\degree} domain boundary.  } 
\label{fig: linear_domain}
\end{figure}

 Here, we derive the width $W_N$ of a N\'eel domain wall induced in the Co overlayer by the exchange bias interaction in the presence of a sharp 180\si{\degree} antiphase AFM domain boundary in the \FeO\ film, and compare this result with the well-known,  analogous calculation for the width $W_B$ of a Bloch wall in the presence of magneto-crystalline anisotropy.  As discussed in the main text, we will assume that the Co magnetisation rotates by 180\si{\degree} at a constant rate throughout the domain wall (figure \ref{fig: linear_domain} \textbf{a}). 
 
As a first step, we will also assume the spin in Co to be co-aligned along the $z$ axis (perpendicular to the film surface).  Assuming the AFM spins in \FeO\ to be aligned along $\pm x$, the magnetisation in the domain wall is described as:
 
 \begin{eqnarray}
 \label{eq: domain_wall_definition}
 M_x&=& M \cos \theta \nonumber\\
 M_y&=& M \sin \theta
 \end{eqnarray}
 
 with
 
 \begin{equation}
 \theta=\pi\left(\frac{x}{W_N}\right)
 \end{equation}
 
 $W_N$ being the \emph{full} width of the domain wall in the $x$ direction.
 
 The non-zero components of the gradient of the normalised magnetisation gradients in Co are:
 
 \begin{eqnarray}
 \frac{\partial m_x}{\partial x}&=& -\pi \frac{\sin \theta}{W_N}\nonumber\\
 \frac{\partial m_y}{\partial x}&=& \pi \frac{\cos \theta}{W_N}
 \end{eqnarray}
 
 The exchange energy is therefore
 
 \begin{eqnarray}
 E_{ex}&=&A_{ex} \int_0^{W_N}  \left( \frac{\partial m_x}{\partial x}\right)^2+\left( \frac{\partial m_y}{\partial x}\right)^2\,dv\nonumber\\
 &=& A_{ex} \pi^2 \frac{\mathcal{A}}{W_N}
 \end{eqnarray}

where $\mathcal{A}=d L$ is the area of the domain wall.  This expression is identical to the exchange energy for a Bloch domain wall in the bulk.

We now calculate the exchange bias energy, subtracting the FM energy as usual.  This results from the following integral over the area:

\begin{equation}
\label{eq: energy_per_area}
E_{ex.b}=\int d\sigma \, K_{ex.b} (1- \cos \theta)
\end{equation}

Where $K_{ex.b}$ is an energy per unit area.  Performing the integral explicitly

\begin{eqnarray}
E_{ex.b}&=&K_{ex.b} L  \; 2 \int_0^{W_N/2} dx \left(1- \cos   \pi \frac{x}{W_N}  \right)\nonumber\\
&=&  K_{ex.b} W_N L \left(1-\frac{2}{\pi}\right)
\end{eqnarray}

Once again, this expression is very similar to the magnetocrystalline anisotropy energy for a Bloch domain wall, with the \emph{caveat} that $K_{ex.b}$ is an energy per unit area, while $K_{an}$ is an energy per unit volume:

\begin{eqnarray}
E_{an}&=&K_{an} L d  \int_0^{W_B} dx \cos^2 \pi \frac{x}{W_B} \nonumber\\
&=&  \frac{1}{2} K_{an} W_B L d 
\end{eqnarray}
 
 By minimising the total energy vs the width of the domain walls, one can easily find:
 
 \begin{eqnarray}
 \label{eq: domain_widths}
 W_N&=&\pi \left(1-\frac{2}{\pi}\right)^{-1/2} \sqrt{\frac{A_{ex} d}{K_{ex.b}}} \approx 5.52  \, L_{ex.b} \nonumber\\
 W_B&=&\pi \sqrt{2}\sqrt{\frac{A_{ex}}{K_{an}}} \approx 4.44 \,L_{an}
\end{eqnarray}

which is consistent with the discussion in Section \ref{section: Analytical considerations} and the definition of the `exchange bias length' in eq. \ref{eq: ex_l_def}.

Relaxing the assumption that the spin in Co to be co-aligned along the $z$ axis, one can let the width of the domain wall depend on $z$, such that:

\begin{equation}
W_N(z)=W_N^0+\lambda z +\dots...
\end{equation}

where the $z$ axis originates at the interface and $\lambda$ is a parameter to be determined by minimising the total energy.  This problem is slightly more complex but is tractable analytically.  To first order, one finds that the Co spins remain strictly co-aligned  (i.e., $\lambda=0$) unless $\frac{d}{L_{ex.b}}\approx 1$, which the case for the 8 \si{\nano\meter} Co film discussed in section \ref{sec: scaling}.

For the purpose of comparing with our simulations, it is also useful to calculate the width of a 60\si{\degree} domain wall (figure \ref{fig: linear_domain} \textbf{b}), which is defined by eq. \ref{eq: domain_wall_definition} together with:

 \begin{equation}
 \theta=\frac{\pi}{3}\left(\frac{x}{W_N}+1\right)
 \end{equation}
 
A very similar calculation to eq. \ref{eq: domain_widths} yields:

\begin{equation}
 \label{eq: domain_widths_60}
 W_N^{60^{\circ}}=\frac{\pi}{3} \left(1-\frac{3}{\pi}\right)^{-1/2} \sqrt{\frac{A_{ex} d}{K_{ex.b}}} \approx 4.93  \, L_{ex.b} 
 \end{equation}

The full width at half maximum is

\begin{equation}
FWHM^{60^{\circ}}=\frac{6}{\pi} \arcsin{\left(\frac{1}{4}\right)} \, W_N^{60^{\circ}} \approx 2.38  \, L_{ex.b}
\end{equation}

\section{Detailed calculation for the analytical merons}
\label{sec: meron_calculations}

Our aim here is to calculate the exchange energy difference between a meron of radius $R$ and a flat vortex with $R \rightarrow 0$, which is expected to be \emph{negative}, since spins in the meron are almost parallel near the core.  We will first discuss the simplest case of $\mathcal{F}(x)=x$, (the `projective' meron).  From eq. \ref{eq: meron_spins} we have

\begin{eqnarray}
m_x&=&-\frac{r}{\sqrt{r^2+R^2}} \sin \phi \nonumber\\
m_y&=& \frac{r}{\sqrt{r^2+R^2}} \cos \phi \nonumber\\
m_z&=& \frac{R}{\sqrt{r^2+R^2}} 
\end{eqnarray}

We will also consider the `linear meron' case:

\begin{eqnarray}
m_x&=&-\sin \theta \sin \phi \nonumber\\
m_y&=& \sin \theta \cos \phi \nonumber\\
m_z&=& \cos \theta 
\end{eqnarray}

where

\begin{equation}
\theta = \left[\begin{array}{c} \frac{\pi r}{2 R} \, \textrm{for}\, r \le R\\\frac{\pi}{2}\, \textrm{for}\, r > R\end{array}\right.
\end{equation}
Using the expression for the gradient in cylindrical coordinates we can easily calculate the exchange energy.  For example for the `projective' meron, 

\begin{eqnarray}
E_{ex}&=&A_{ex} \int dv \left(\frac{\partial}{\partial r}\right)^2+\left(\frac{1}{r}\frac{\partial}{\partial \phi}\right)^2\nonumber\\
&=&2 \pi A_{ex} d \int r \, dr  \frac{r^2+2R^2}{(r^2+R^2)^2}\nonumber\\
&=& \left .\pi A_{ex} d \left(\ln (r^2+R^2)-\frac{R^2}{r^2+R^2} \right)\right \vert_0^\infty\nonumber\\
&=&\pi  A_{ex} d \left(2 \lim_{r \rightarrow \infty} \ln \left(\frac{r}{R}\right) +1 \right)
\end{eqnarray}

The general expression

\begin{equation}
E_{ex}=\pi  A_{ex} d \left(2 \lim_{r \rightarrow \infty} \ln \left(\frac{r}{R}\right) +c \right)
\end{equation}

holds in a variety of situations --- in particular, when  $\mathcal{F}(x)=x^n$ is a positive power of $x$, one can show that c=n.  Moreover, if $\mathcal{F}(x)$ is \emph{zero} outside a radius $R$ and $R$ is the only length-scale involved, then $E_{ex}$ must be independent on $R$ due to simple dimensional considerations.

From $E_{ex}$, we must subtract the energy of a planar vortex ($R=0$), where it is convenient to replace the lower limit of integration with a small length $a$, which will be sent to zero at the end of the calculation.  The vortex energy integrated to infinity is

\begin{equation}
 E_{vortex}^{\infty} = \pi  A_{ex} d \left(2 \lim_{r \rightarrow \infty} \ln \left(\frac{r}{a}\right) \right)
\end{equation}

while the vortex energy integrated to a radius $R$ is:

\begin{equation}
 E_{vortex}^{R} = \pi  A_{ex} d \left(2  \ln \left(\frac{R}{a}\right) \right)
\end{equation}

By performing the subtraction, one obtains the following general formula for the pure exchange energy of the core:

\begin{equation}
\label{eq: Energy core}
\Delta E_{ex} =\pi A_{ex} d  \left(-2 \ln \left(\frac{R}{a}\right) + c\right)
\end{equation}

which is always \emph{negative} for $R \gg a$,  and

\begin{equation}
\label{eq: Energy core diff}
\frac{\partial \Delta E_{ex}}{\partial R}=-2\pi A_{ex} d \frac{1}{R}
\end{equation}

We now need to calculate the loss of exchange bias energy occurring at the interface with respect to the vortex, due to the out-of-plane canting, which is obtained by performing the surface integral in eq. \ref{eq: energy_per_area}.  With a straightforward calculation one obtains for the projective meron ($\mathcal{F}(x)=x$):

\begin{equation}
\label{eq: ex_b projective}
\Delta E_{ex.b}=\frac{\pi}{2} K_{ex.b} R^2\left( \ln \frac{4 r^2}{R^2}-1 \right)
\end{equation}

which has a logarithmic divergence, due to the fact that the $m_z$ does not decay fast enough away from the core, while for the quadratic meron ($\mathcal{F}(x)=x^2$):

\begin{equation}
\label{eq: ex_b quadratic}
\Delta E_{ex.b}=\pi K_{ex.b} R^2
\end{equation}

and

\begin{equation}
\label{eq: ex_b quadratic diff}
\frac{\partial \Delta E_{ex.b}}{\partial R}=2\pi K_{ex.b} R
\end{equation}

For the linear meron, the equivalent expressions are:

\begin{equation}
\label{eq: ex_b linear}
\Delta E_{ex.b}=2 \pi K_{ex.b} \int_0^R r \sin \left(\frac{\pi r}{2 R}\right) dr=\frac{8}{\pi}K_{ex.b} R^2
\end{equation}

and

\begin{equation}
\label{eq: ex_b linear diff}
\frac{\partial \Delta E_{ex.b}}{\partial R}=2\pi \frac{8}{\pi^2}K_{ex.b} R
\end{equation}

We need to minimise the expression

\begin{equation}
\Delta E_{tot}=\Delta E_{ex.b}+\Delta E_{ex}
\end{equation}

as a function of R, which is easily done with the help of equations \ref{eq: Energy core diff}, \ref{eq: ex_b quadratic diff} and \ref{eq: ex_b linear diff}, yielding:

\begin{equation}
\label{eq: meron core width}
R_{min}=\kappa \sqrt{ \frac{A_{ex}d}{K_{ex.b} }}=\kappa L_{ex.b}
\end{equation}  

where $\kappa=1$ for the quadratic meron and $\kappa=\pi \sqrt{2}/4 \approx 1.11$ for the linear meron.

The linear meron can be directly compared with the linear N\'eel domain wall by observing that the lengths over which the spins rotate by 180$^{\circ}$ are $W_N\approx 5.52 L_{ex.b}$ (eq. \ref{eq: domain_widths}) and $W_{core}=2 R_{min}\approx 2.22  L_{ex.b}$ (eq. \ref{eq: meron core width}).  Another useful quantity is the FWHM of the $M_z$ peak, which is directly comparable to our simulations.  A very simple analysis yields:

\begin{equation}
\label{eq: fwhm_core}
FWHM=\frac{2}{3}W_{core} \approx 1.5 L_{ex.b}
\end{equation}

\bibliography{../Fe2O3_vortices.bib}

\end{document}